\documentclass[aps,prl,notitlepage,twocolumn,superscriptaddress,longbibliography,nofootinbib]{revtex4-1}
\pdfoutput=1
\usepackage[utf8]{inputenc}

\usepackage{amsmath,amsthm,amssymb,amsfonts}
\usepackage{scalerel}
\usepackage{mathtools}
\usepackage{graphicx}
\usepackage{stfloats}
\usepackage{subfigure}
\usepackage[normalem]{ulem}
\usepackage[colorlinks = true,
            linkcolor = blue,
            urlcolor  = blue,
            citecolor = blue,
            anchorcolor = blue]{hyperref}
\usepackage{mathrsfs}
\usepackage{bbold}
\usepackage{units}
\allowdisplaybreaks
\usepackage{bm}
\usepackage{braket}
\usepackage{makecell}
\usepackage[nameinlink]{cleveref} 
\usepackage{upgreek}
\usepackage{blindtext}
\usepackage{verbatim}
\usepackage{algorithm}
\usepackage[noend]{algpseudocode}
\usepackage[dvipsnames]{xcolor}
\usepackage{bbm}
\usepackage{array}
\usepackage{multirow}
\usepackage{tabularx}
\usepackage{float}
\usepackage{dcolumn}
\usepackage{slashed}
\usepackage{braket}
\usepackage{verbatim}
\usepackage{multirow}
\usepackage{resizegather}
\usepackage{titlesec}
\usepackage[toc,page]{appendix}
\usepackage[export]{adjustbox}

\newcommand{\tr}{\mathrm{tr}}
\theoremstyle{definition}
\newtheorem{thm}{Theorem}

\newtheorem*{prop*}{Property}

\newtheorem{ctr}{Conjecture}
\newtheorem*{ctr*}{Conjecture}
\newtheorem{lemma}{Lemma}

\begin{document}

\title{Mixed-State Entanglement in a Minimal Model of Quantum Chaos\\
}
\author{Tanay Pathak\,\,\href{https://orcid.org/0000-0003-0419-2583}
{\includegraphics[scale=0.05]{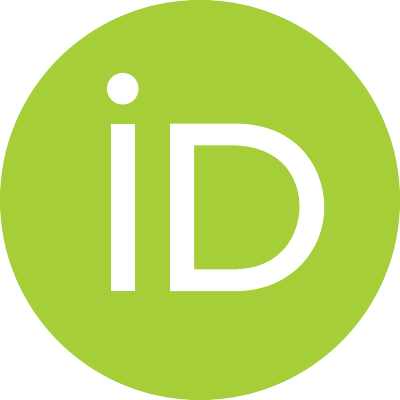}}}
\email{pathak.tanay.4s@kyoto-u.ac.jp}
\affiliation{Department of Physics, Kyoto University, Kitashirakawa Oiwakecho, Sakyo-ku, Kyoto 606-8502, Japan}

\begin{abstract}
Understanding the dynamics of quantum correlations in many-body systems is a central problem in non-equilibrium quantum physics. We study the spread of mixed-state entanglement in a minimal model of quantum chaos, the kicked field Ising model.
By combining the replica trick with the space-time duality of the model, we determine the exact spectrum of the partially transposed reduced density matrix. The resulting flat spectrum leads to exact relations between entanglement negativity, odd entropy and R\'enyi mutual information at early times. Numerical results further demonstrate that for equal tri-partitions and at late times, all entanglement measures saturate to the Haar-random values. In contrast, for unequal tri-partitions R\'enyi mutual information and negativity vanish at late times, implying that the corresponding reduced density matrix is factorizable. Extensive numerical simulations also show that the relation remains quantitatively valid for generic initial states, leading us to conjecture it for all initial states and all times.
\end{abstract}
\maketitle


\textit{Introduction---}Analytical understanding of phenomena in quantum many-body systems is quite challenging due to its exponentially large Hilbert space. Due to this the task of studying time evolving observables is complicated, even numerically. Recently, a new approach to study quantum many body systems was put forth that utilizes the discrete space time duality to understand the dynamics \cite{PhysRevLett.102.240603,Akila_2016}(see \cite{Bertini:2025ddr} for a review). This has been quite fruitful and used to study various phenomena in many body physics such as exact dynamics of correlation functions \cite{PhysRevLett.123.210601,PhysRevLett.126.100603,PhysRevB.102.174307,PhysRevB.101.094304,PhysRevLett.133.170403}, spread of entanglement and related properties \cite{Bertini:2018fbz,PhysRevLett.132.120402,  Gopalakrishnan:2019pip,Bertini:2019gbu,Bertini:2019wkb,PhysRevLett.125.070501,Reid:2021fsg,Zhou:2022uuw,PhysRevB.107.174311,PRXQuantum.6.010324}, spectral form factor \cite{Bertini:2018wlu,Bertini:2020mdd,PhysRevResearch.2.043403,PhysRevX.11.021051,PhysRevResearch.6.033226,PhysRevB.111.094316} amongst many.
Within this framework, the dual unitary circuits have emerged as an important class of exactly solvable models and had been studied quite well. The kicked field Ising model (KFIM) \cite{prosenkfim,PhysRevE.65.036208,Prosen:2007hwp} further presents itself as a special class of these dual unitary circuits. Owing to the dual unitarity property \cite{Akila_2016}, the model serves as a minimal setting for studying quantum chaos and entanglement dynamics in interacting many-body systems.

In this paper we study the spread of entanglement, in a class of solvable states, in general setting of mixed state entanglement. In particular we are interested in the entanglement between regions $A$ and $B$ of tripartition $ABC$ of a time evolved initial state. For generic one-dimensional systems with discrete space-time structure and local interactions, a universal relation between the entanglement negativity \cite{Calabrese:2012ew,PhysRevA.65.032314,Kudler-Flam:2018qjo} and the R\'enyi-$1/2$ mutual information \cite{Wolf:2007tdq,Alba:2018hie} at early times, was derived in \cite{PhysRevLett.129.140503} generalizing various other previous results between these quantities \cite{Alba:2018hie,Gruber:2020afu,PhysRevB.102.235110,Murciano:2021zvs} . Here we show that, for a class of initial solvable states, this relation is satisfied in the KFIM at its dual point, providing an independent proof of the universal result in this minimal model of quantum chaos. Moreover, we show that the spectrum of partially transposed reduced density matrix involved is flat which in turn fixes the mixed state correlations. This also allow us to evaluate exactly, another mixed state entanglement measure, the odd entropy \cite{PhysRevLett.122.141601,Mollabashi:2020ifv,Kudler-Flam:2020url}. Finally, with extensive numerics we also show that such a relation is also valid for generic states and at late times as well. It is also revealed that if the three subsystem sizes are unequal then at late times the density matrix $\rho_{AB}$ of subsystem $AB$ is factorizable as indicated by vanishing mutual information (and negativity). The odd entropy is non-zero and equal to the von Neumann entanglement entropy of $AB$, in agreement with \cite{PhysRevLett.122.141601}. For the case of generic state, not lying in the solvable class, we numerically demonstrate this feature. 

\textit{Setup and results---}
We consider the KFIM, for which the Hamiltonian is given by \cite{prosenkfim,PhysRevE.65.036208,Prosen:2007hwp} 

\begin{align}\label{eq:hamxyz}
    H &= H_{I} + H_{K} \sum_{\tau= -\infty}^{\infty} \delta(t- \tau) 
     \\
 \text{where},\, H_{I} & = J \sum_{i=1}^{L}  \sigma^{z}_{i}\sigma^{z}_{i+1} + \sum_{i=1}^{L}h_{i}\sigma_{i}^{z}, \nonumber \\
    H_{K}& = \sum_{i=1}^{L} b\, \sigma^{x}_{i}.\nonumber
\end{align}
The total Floquet operator of the system is 
\begin{equation}
    U= U_{K}U_{I},
\end{equation}
where $U_{I}= e^{-i H_{I}}$ and $U_{K}= e^{-i H_{K}}$. In this paper we will be only interested in the behavior at the dual unitary given by $|J|= |b|= \frac{\pi}{4}$ and  $h_{i}$ can be generic. We specifically consider $J= \frac{\pi}{4}, b= -\frac{\pi}{4}$ for our analysis. 

We consider the initial state, which is a product state, as follows:
\begin{equation}
    \ket{\psi_{\theta,\phi}}= \bigotimes_{k=1}^{L} \,\left(\cos\left(\frac{\theta_{k}}{2}\right) \ket{\uparrow} + e^{i\phi_{k}} \sin\left(\frac{\theta_{k}}{2}\right) \ket{\downarrow}\right).
\end{equation}
To study the entanglement dynamics we specifically consider the two classes of states: transverse ($\mathcal{T}$) and longitudinal ($\mathcal{L}$), which are given by following parameters
\begin{align}
    &\mathcal{T}= \{\ket{\psi_{\pi/2,\phi}} \forall\, \theta_{k}= \frac{\pi}{2}, \phi \in [0,2\pi] \},  \\
    &\mathcal{L}= \{\ket{\psi_{\bar{\theta},\phi}} \bar{\theta}=\{ 0,\pi\}, \phi_{k} \in [0,2\pi]\}.
\end{align}
We refer to above states as solvable in the sense that the dynamics can be calculated exactly if the initial state belong to either $\mathcal{T}$ or $\mathcal{L}$ class \cite{Bertini:2018fbz}. States which do not belong to these class will henceforth be called generic. 

In the following we now consider a tri-partition of state $ABC$ with subsystem size $L_{A},L_{B}$ and $L_{C}$. We are then interested in the dynamics of entanglement between regions $A$ and $B$ using  entanglement negativity
\begin{align}\label{eq:negativity}
    \mathcal{E}(t)= \ln \tr\left(\sqrt{(\rho_{AB}^{T_{B}}(t))^{\dagger}\rho_{AB}^{T_{B}}(t)}\right),
\end{align}
where $\rho_{AB}(t)= \tr_{C}(\ket{\psi(t)}\bra{\psi(t)}$ is the reduced density matrix of subsystem $AB$ at time $t$ and $(\cdot)^{T_{B}}$ denotes the partial transpose with respect to $B$. 
To evaluate the entanglement negativity we setup a replica trick to obtain the even moments of the partial transposed reduced density matrix 
\begin{equation}\label{eq:moments}
    \mathcal{E}_{2n}(t)= \tr((\rho_{AB}^{T_{B}}(t))^{2n}).
\end{equation}
The even moments are given by the following lemma.
\begin{lemma}\label{eq:lemma}
For large $L_A,L_B, L_{C}$ and the initial state belonging to $\mathcal{T}$ class, the even moments $\mathcal{E}_{2n}(t)$ are given as follows
    \begin{equation*}
  \mathcal{E}_{2n}(t)=  \ln( \tr((\rho_{AB}^{T_{B}}(t)^{2n})))= (4-6n)t\ln(2).
\end{equation*}
\end{lemma}
The above result holds in general for early times i.e. $2t \leq L_{A}, L_{B}, L_{C}$. Entanglement negativity, Eq. \eqref{eq:negativity} can be recovered using the above lemma by analytically continuing $2n \rightarrow \alpha$ and then taking the limit $\alpha \rightarrow 1$. Finally, we obtain
\begin{align}
    \mathcal{E}(t)= \ln(2)t.
\end{align}

Then not that for a tri-partition $ABC$ the R\'enyi mutual information between partitions $A$ and $B$ (after $C$ is traced out) is given as follows
\begin{equation}
    I_{A:B}^{(\alpha)}(t)= S^{(\alpha)}_{A}(t)+S^{(\alpha)}_{B}(t)- S^{(\alpha)}_{AB}(t),
\end{equation}
where $S^{(\alpha)}_{A}= \frac{1}{1-\alpha}\log(\tr(\rho_{A}^{\alpha}))$ denotes the R\'enyi entropy with $\rho_{A}$ being the reduced density matrix of subsystem $A$. At the dual unitary point using previous results on R\'enyi entropy \cite{Bertini:2018fbz} we obtain
\begin{align}\label{eq:mutualinfo}
    I^{(\alpha)}_{A:B}(t)&= \nonumber \\
    &(\min(2t,L_{A})+ \min(2t,L_{B})- \min(2t,L_{AB}))\ln(2).
\end{align}

Combining \eqref{eq:negativity} and \eqref{eq:mutualinfo} gives us the following for the early times 
\begin{equation}\label{eq:fundrelation1}
   2 \mathcal{E}(t) = I_{A:B}^{(\alpha)}(t)= S_{A}^{(\alpha)}(t)= S^{(\alpha)}_{B}(t).
\end{equation}

Furthermore, for large $L_{A}, L_{B}, L_{C}$, we have following theorem for the spectrum (singular values or absolute value of eigenvalues) of $\rho^{T_{B}}_{AB}(t)$, denoted by $\textrm{Spec}[\rho_{AB}^{T_{B}}(t)] $.
\begin{thm} \label{theorem1}
The spectrum of $(\rho_{AB}^{T_{B}}(t))$, corresponding to $\mathcal{T}$ type initial state, for early times, has following properties--
\begin{enumerate}
   \item $\textrm{Spec}[\rho_{AB}^{T_{B}}(t)]= \{2^{-3t},0\}$
with degeneracy of non-zero singular value equals to $2^{4t}$. 
\item The number of positive and negative eigenvalues which are given as follows 
  \begin{equation*}
     N_{+}= \frac{2^{4t}+2^{3t}}{2},\, N_{-}= \frac{2^{4t}-2^{3t}}{2}.
\end{equation*}
\end{enumerate}
\end{thm}
See \cite{supp} for proof of these properties. To be noted that the presence of negative eigenvalues in the spectrum of partially transposed reduced density matrix is an indication of the presence of quantum correlations in the system \cite{PhysRevLett.77.1413}. Although stated for large subsystem sizes, the Theorem \ref{theorem1} is holds for early times i.e. $ 2t \leq L_{A},L_{B},L_{C} $. 

Using these properties it is then straightforward to obtain another measure of mixed state entanglement, the odd entropy \cite{PhysRevLett.122.141601} which is defined as 
\begin{equation}
    \mathcal{E}^{(o)}(t)= -\sum_{\lambda_{i}>0}|\lambda_{i}|\log(|\lambda_{i}|)+ \sum_{\lambda_{i}<0}|\lambda_{i}|\log(|\lambda_{i}|),
\end{equation}
where $\lambda_{i}$ are the eigenvalues of $\rho_{AB}^{T_{B}}(t)$. Using Theorem \eqref{theorem1} the odd entropy evaluates to
\begin{equation}
    \mathcal{E}^{(o)}(t)= 3\log(2)t.
\end{equation}
Combining with Eq. \eqref{eq:fundrelation1} we obtain
\begin{equation}\label{eq:fundrelation2}
     2\mathcal{E}(t) = \frac{2}{3}\mathcal{E}^{o}(t) = I_{A:B}^{(\alpha)}(t) = S^{(\alpha)}_{A}(t)= S^{(\alpha)}_{B}(t).
\end{equation}
Notice that due to the flat spectrum of the model the identity is valid for general $\alpha-$ R\'enyi mutual information and $\alpha-$ R\'enyi entropies. Similar results can also be obtained for $\mathcal{L}$ class as well by noting that its results are delayed by a period as compared to $\mathcal{T}$ class.
\begin{figure}[htbp]
\centering
    \includegraphics[width= \linewidth]{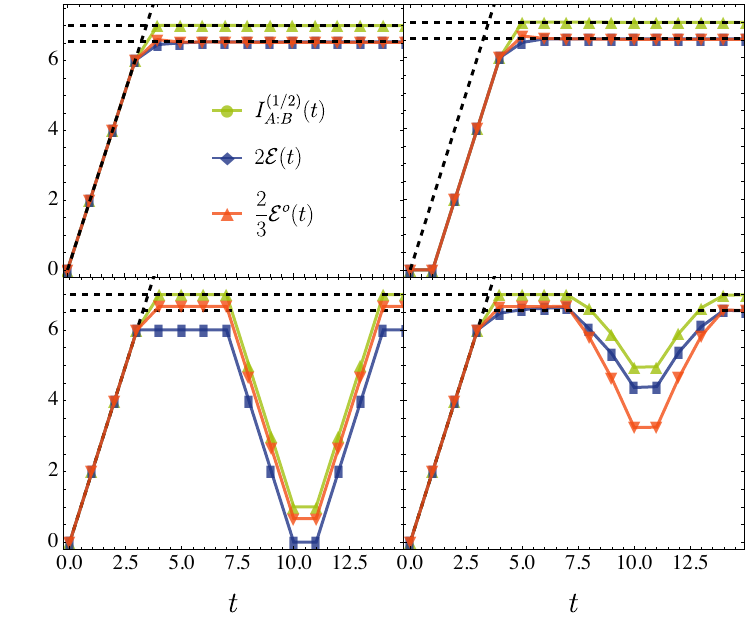}
    \caption{ R\'enyi-1/2 mutual information ($I_{A:B}^{(1/2)}(t)$), negativity ($2\mathcal{E}(t)$) and odd entropy($\frac{2}{3}\mathcal{E}^{o}(t)$) for initial solvable state. All the systems have total size $L=21$ and subsystem size $L_{A}=L_{B}=L_{C}=7$. We choose the parameters corresponding to: (a) $\mathcal{T}$ class: $\theta=\pi/2$, $\phi=0$,$h_{i}=1$. (b) $\mathcal{L}$ class: $\theta=0$,$\phi=0$,$h_{i}=1$. (c) Integrable: $\theta=\pi/2,\phi=0,h_{i}=0$ and (d) Weak integrability breaking: $\theta=\pi/2,\phi=0,h_{i}=0.1.$ The two horizontal black dashed lines denote the Haar random values for negativity (lower) and mutual information (upper).}\label{fig:sminfonegativeallf}
\end{figure}

\begin{figure}[tbp]
    \centering
    \includegraphics[width= \linewidth]{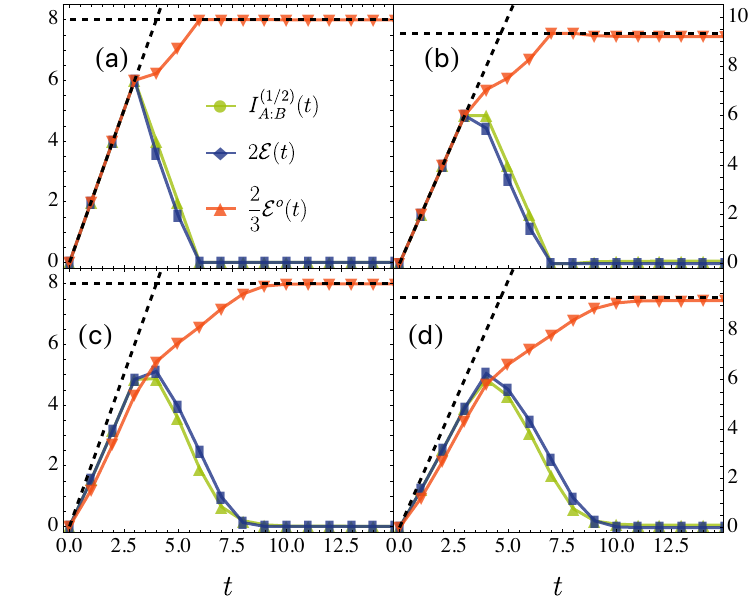}
    \caption{R\'enyi-$1/2$ mutual information ($I_{A:B}^{(1/2)}(t)$), negativity ($2\mathcal{E}(t)$) and odd entropy($\frac{2}{3}\mathcal{E}^{o}(t)$) for initial state corresponding to $\mathcal{T}$class in (a) and (b) and generic state with $\theta=\phi=1$ in (c) and (d). All the systems have total size $L=30$. The subsystem sizes are (a) $L_{A}=L_{B}=6$, (b) $L_{A}=L_{B}=7$, (c) $L_{A}=L_{B}=6$, (d) $L_{A}=L_{B}=7$, and $L_{C}=L-(L_{A}+L_{B})$. The horizontal black dashed lines denotes the value $\frac{2}{3}S_{AB}$.}\label{fig:minfonegativeallf}
\end{figure}

Next, we numerically confirm the derived results and also report their validity for finite size systems. We take total system size of $L=21$ and subsystem sizes as $L_{A}=L_{B}=L_{C}=7$.
In Fig. \eqref{fig:sminfonegativeallf} we show results for solvable state the transverse and the longitudinal state. Also shown are the results obtained for integrable ($h_{i}=0$) and system with weak integrability breaking ($h_{i}=0.1$). The \emph{linear growth regime} follows the relation \ref{eq:fundrelation2} exactly. In the saturation regime, however, we find differences, which we attribute to the finite size effect and expect it to go to zero for larger system sizes. For late times $2t \geq L_{A},\,L_{B},\,L_{C}$, the values are found to be numerically equal to the value of the quantities corresponding to a Haar random state.
\begin{figure}[htbp]
    \centering
    \includegraphics[width= \linewidth]{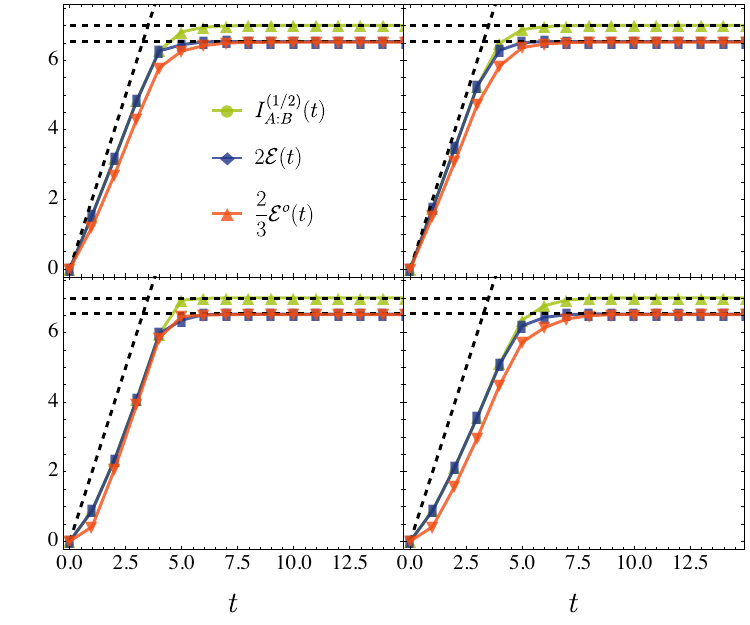}
    \caption{R\'enyi-1/2 mutual information ($I_{A:B}^{(1/2)}(t)$), negativity ($2\mathcal{E}(t)$) and odd entropy($\frac{2}{3}\mathcal{E}^{o}(t)$) for generic intial state. All the systems have total size $L=21$, subsystem size $L_{A}=L_{B}=L_{C}=7$ and $h_{i}=1$. We choose the parameters corresponding to: (a) $\theta=1,\phi=1$. (b) $\theta=2,\phi=2$ (c) $\theta=2.5,\phi=1$(d) $\theta=2.5,\phi=3$. The two horizontal black dashed lines denote the Haar random value for negativity (lower) and mutual information (upper).}\label{fig:nsminfonegativeallf}
\end{figure}
In Fig. \eqref{fig:nsminfonegativeallf} we consider initial states which are generic. Similar to the case of solvable states we observe that the relation \eqref{eq:fundrelation2} holds at early times and the late time saturation agrees with the Haar random value. These results extend the validity of the results for generic initial states which are not necessarily solvable. Next, we consider the case of unequal partition, for both solvable and generic initial states, in Fig. \eqref{fig:minfonegativeallf}. For this case we consider $L=28$ and subsystem sizes as $L_{A}=L_{B}=6,7$ and $L_{C}=L-(L_{A}+L_{B})$. The initial states belong to solvable $\mathcal{T}$ class and generic state with $\theta=\phi=1$. We find that at late times, entanglement negativity and mutual information go to zero, which implies there is no distillable entanglement. Furthermore, we numerically observe that for this case the spectrum is still \emph{flat}, but $N_{-}=0$, which explains the zero negativity at late times. Based on analytical results and various numerical calculations we are led to the conjecture
\begin{ctr}\label{conjec1}
    \begin{equation}
     2\mathcal{E}(t)  = I_{A:B}^{(\alpha)}(t),
\end{equation}
hold for generic states at all times $t$. Specifically, for $2t > L_{A},L_{B},L_{C}$, we have 
\begin{equation*}
    2\mathcal{E}(t) = I_{A:B}^{(\alpha)}(t)\begin{cases}
        = 0 \quad L_{A}\neq L_{B}\neq L_{C}, \\
        \neq 0 \quad L_{A}=L_{B}=L_{C},
    \end{cases}
\end{equation*}
where the non-zero late time value is equal to Haar random value. 
\end{ctr}
For the case of equal subsystem sizes vanishing of mutual information implies that at late times the density matrix, $\rho_{AB}$ factorizes \cite{Witten:2018zva} i.e. $\rho_{AB}= \rho_{A} \otimes \rho_{B}$.
This behavior is similar to the Page curve \cite{PhysRevLett.71.3743,Page:2013dx,RevModPhys.93.035002}, for mixed state entanglement, which is of relevance for black hole information paradox. Interestingly, for the case of unequal subsystem sizes we obtain non-zero odd entropy at late times, Fig. \eqref{fig:minfonegativeallf}. In fact, we verify numerically that at late times, the odd entropy is equal to von Neumann entropy of subsystems $AB$ (with appropriate finite size effects)
\begin{equation}\label{eq:oentropyrel}
    \lim_{2t > L_{A},L_{B},L_{C} }\mathcal{E}^{(o)}(t)= S^{(vN)}_{AB}(t)
\end{equation}

This observation is in agreement with the property that, if $\rho_{AB}$ is a product state then the odd entropy reduces to the von Neumann entanglement entropy \cite{PhysRevLett.122.141601}. Preliminary numerical tests also suggest the validity of Conjecture \ref{conjec1} away from the dual unitary point as well, where we consider perturbed parameters: $J= \frac{\pi}{4}-\epsilon$ and $b= -\frac{\pi}{4}-\epsilon$.

\textit{Discussions---}In this paper we establish a relation between mixed-state entanglement measures for a class of solvable states evolved using KFIM. Setting up the replica trick we determine the exact spectrum of the partially transposed reduced density matrix at early times. This allowed us to relate entanglement negativity, odd entropy and R\'enyi mutual information. Furthermore, precise numerical calculations suggest that the relation between entanglement negativity and R\'enyi mutual information persists at late times. In contrast, the behavior of odd entropy depends on subsystem geometry. For equal partitions Eq. \eqref{eq:fundrelation2} continues to hold, whereas for $L_C > L_A, L_B$ the odd entropy reduces to the von Neumann entanglement entropy of subsystem $AB$. In this regime the mutual information (along with negativity) vanishes, implying that the reduced density matrix factorizes.

While our analytical results are derived for the self-dual KFIM, the general result for moments given by Lemma \ref{eq:lemma} and the corresponding dual transfer matrix involved (see \cite{supp} for details) suggests that the relation \eqref{eq:fundrelation2} is also valid in a broader class of many-body models with dual-unitary point. Preliminary calculations also show that the relation \eqref{eq:fundrelation2} can be obtained for kicked chain with local dimensions $q>2$ \cite{PhysRevB.102.174307,PhysRevB.105.144306,Miao:2023mcr,Claeys:2024tuy} and for other classes of generalized kicked Ising chains \cite{PhysRevResearch.7.L012011,PhysRevLett.132.120402,PhysRevE.110.L022101,Bertini:2025ddr}. It would also be interesting to consider disjoint regions $A$ and $B$ to probe the entanglement dynamics in a more refined manner. Numerically, we also find that the relations remain valid under small perturbations away from the dual-unitary point. It is therefore important to gain a better understanding of these numerical results via analytical perturbative methods as well \cite{PhysRevX.11.011022}. The universal relation derived in \cite{PhysRevLett.129.140503} was argued to be valid for local interactions. It would therefore be important to systematically investigate whether this relation persists in systems with long-range interactions or in circuits with nonlocal gates, such as tri-unitary gates \cite{Jonay:2021kgl} and their systematic generalizations, where analytical control is more limited.

\emph{Acknowledgments:}
The author is grateful to Abhik Kumar Saha, Kenya Tasuki and  Masaki Tezuka for various useful suggestions.
The work is supported by JST CREST (Grant No. JPMJCR24I2).
\twocolumngrid
\bibliography{references}


\hbox{}\thispagestyle{empty}\newpage

\onecolumngrid
\raggedbottom
\begin{center}
\textbf{\large Supplemental Material: Mixed-State Entanglement in a Minimal Model of Quantum Chaos}
\vspace{2ex}

{\normalsize Tanay Pathak$^{1}$ \,\,\href{https://orcid.org/0000-0003-0419-2583}
{\includegraphics[scale=0.05]{orcidid.pdf}}}\\
\small $^{1}$Department of Physics, Kyoto University, Kitashirakawa Oiwakecho, Sakyo-ku, Kyoto 606-8502, Japan

\end{center}

\appendix
In this supplemental material, we provide the proofs of Lemma 1 and Theorem 1 of the main text. In particular
\begin{itemize}
    \item In Section \ref{sec:replica} we set up the replica trick for the partially transposed reduced density matrix. 
    \item In Section \ref{sec:theolem1} we prove Lemma 1 and Theorem 1 of the main text. 
\end{itemize}

\section{Entanglement negativity and the replica trick}\label{sec:replica}

To evaluate entanglement negativity, we first setup the replica trick. Consider the initial state as a  product state given as follows

\begin{equation}
    \ket{\psi_{\theta,\phi}}= \bigotimes_{k=1}^{N} \,\left(\cos\left(\frac{\theta_{k}}{2}\right) \ket{\uparrow} + e^{i\phi_{k}} \sin\left(\frac{\theta_{k}}{2}\right) \ket{\downarrow}\right) .
\end{equation}
To study the entanglement dynamics, we choose two classes of states: Transverse ($\mathcal{T}$) and Longitudinal ($\mathcal{L}$), which are given by following parameters
\begin{align}
    &\mathcal{T}= \{\ket{\psi_{\theta_{k},\phi}} \forall\, \theta_{k}= \frac{\pi}{2}, \phi \in [0,2\pi] \},\label{suppeq:tclass} \\
    &\mathcal{L}= \{\ket{\psi_{\bar{\theta},\phi}} \bar{\theta}=\{ 0,\pi\}, \phi_{k} \in [0,2\pi]\}. \label{suppeq:lclass}
\end{align}
Considering tri-partitions $A, B$ and $C$ of time evolved state $\ket{\psi_{\theta,\phi}}$, the corresponding reduced density matrix is given by 
\begin{align}
        \rho_{ABC}= \sum_{i_{1},i_{2},i_{3} \atop i'_{1},i'_{2},i'_{3} } \braket{i_{1},i_{2},i_{3}| U_{KI}[h]^{t}|\psi_{\theta,\phi}} \braket{\psi_{\theta,\phi}| U_{KI}[h]^{-t}|i'_{1},i'_{2},i'_{3}} \times \ket{i_{1},i_{2},i_{3}}\bra{i'_{1},i'_{2},i'_{3}}.
    \end{align}
Tracing out $C$, we obtain
    
    \begin{align*}
        \tr_{C}(\rho_{ABC})= \sum_{k}\sum_{i_{1},i_{2},i_{3} \atop i'_{1},i'_{2},i'_{3} } \braket{i_{1},i_{2},i_{3}| U_{KI}[h]^{t}|\psi_{\theta,\phi}} \braket{\psi_{\theta,\phi}| U_{KI}[h]^{-t}|i'_{1},i'_{2},i'_{3}} \times (\mathbb{I}_{AB}\otimes \bra{k})\ket{i_{1},i_{2},i_{3}}\bra{i'_{1},i'_{2},i'_{3}}(\mathbb{I}_{AB}\otimes \ket{k}),
    \end{align*}
which after simplification gives us
 \begin{align}
       \rho_{AB}\equiv \tr_{C}(\rho_{ABC})= \sum_{i_{1},i_{2},i_{3} \atop i'_{1},i'_{2}} \braket{i_{1},i_{2},i_{3}| U_{KI}[h]^{t}|\psi_{\theta,\phi}} \times \braket{\psi_{\theta,\phi}| U_{KI}[h]^{-t}|i'_{1},i'_{2},i_{3}} \times \ket{i_{1},i_{2}}\bra{i'_{1},i'_{2}}.
    \end{align}
Next, taking the partial transpose with respect to system $B$, we obtain
 \begin{align}
       & \rho_{AB}^{T_{B}}= \sum_{a_{1},b_{1},c_{1} \atop a_{2},b_{2}} \braket{a_{1},b_{1},c_{1}| U_{KI}[h]^{t}|\psi_{\theta,\phi}}  \times \braket{\psi_{\theta,\phi}| U_{KI}[h]^{-t}|a_{2},b_{2},c_{1}} \times \ket{a_{1},b_{2}}\bra{a_{2},b_{1}}.   
       \end{align}
We can then obtain the even moments as
     \begin{align}
       & \tr((\rho_{AB}^{T_{B}})^{2n})= \sum_{k_{1},k_{2}}\sum_{a_{1},b_{1},c_{1} \atop a_{2},b_{2}} \braket{a_{1},b_{1},c_{1}| U_{KI}[h]^{t}|\psi_{\theta,\phi}} \times \braket{\psi_{\theta,\phi}| U_{KI}[h]^{-t}|a_{2},b_{2},c_{1}} \times  \nonumber \\
       &\times \sum_{a_{3},b_{3},c_{2} \atop a_{4},b_{4}} \braket{a_{3},b_{3},c_{2}| U_{KI}[h]^{t}|\psi_{\theta,\phi}} \times \braket{\psi_{\theta,\phi}| U_{KI}[h]^{-t}|a_{4},b_{4},c_{2}} \nonumber \\
       \vdots \nonumber \\
    &\times \sum_{a_{4n-1},b_{4n-1},c_{2n} \atop a_{4n},b_{4n}} \braket{a_{4n-1},b_{4n-1},c_{2n}| U_{KI}[h]^{t}|\psi_{\theta,\phi}} \times \braket{\psi_{\theta,\phi}| U_{KI}[h]^{-t}|a_{4n},b_{4n},c_{2n}}  \nonumber \\
& \braket{k_{1},k_{2}|a_{1},b_{2}}\braket{a_{2},b_{1}|a_{3},b_{4}} \braket{a_{4},b_{3}|a_5,b_6} \cdots \braket{a_{4n-2},b_{4n-3}|a_{4n-1},b_{4n}}\braket{a_{4n},b_{4n-1}|k_{1},k_{2}}.
    \end{align}

Summing over $k_{1}, k_{2}$ we obtain the condition on indices as: $a_{4n}=a_{1}$, $a_{3}=a_{2}$, $a_{5}=a_{4}\cdots$  $a_{4n-2}=a_{4n-1}$
and $b_{1}=b_{4},b_{2}=b_{4n-1}$, $b_{3}=b_{6}\cdots$  $b_{4n-3}= b_{4n}$. Next, we relabel the indices as $b_{2j} \rightarrow \tilde{b}_{j}$ and obtain

     \begin{align}
       & \tr((\rho_{AB}^{T_{B}})^{2n})= \sum_{\{a_{i}\},\{\tilde{b}_{i}\},\{c_{i}\}} \braket{a_{1},\tilde{b}_{2},c_{1}| U_{KI}[h]^{t}|\psi_{\theta,\phi}} \times \braket{\psi_{\theta,\phi}| U_{KI}[h]^{-t}|a_{2},\tilde{b}_{1},c_{1}}  \nonumber \\
       &\times  \braket{a_{2},\tilde{b}_{3},c_{2}| U_{KI}[h]^{t}|\psi_{\theta,\phi}} \times \braket{\psi_{\theta,\phi}| U_{KI}[h]^{-t}|a_{3},\tilde{b}_{2},c_{2}} \nonumber \\
       \vdots \nonumber \\
    &\times  \braket{a_{2n},\tilde{b}_{1},c_{2n}| U_{KI}[h]^{t}|\psi_{\theta,\phi}} \times \braket{\psi_{\theta,\phi}| U_{KI}[h]^{-t}|a_{1},\tilde{b}_{2n},c_{2n}}.  \nonumber \\
    \end{align}
Finally, renaming all the $2n-$ indices we obtain

     \begin{align}\label{eq:evenmomsf}
       & \tr((\rho_{AB}^{T_{B}})^{2n})= \sum_{\{a_{i}\},\{b_{i}\},\{c_{i}\}} \braket{a_{1},b_{2},c_{1}| U_{KI}[h]^{t}|\psi_{\theta,\phi}} \times \braket{\psi_{\theta,\phi}| U_{KI}[h]^{-t}|a_{2},b_{1},c_{1}} \times  \nonumber \\
       &\times  \braket{a_{2},b_{3},c_{2}| U_{KI}[h]^{t}|\psi_{\theta,\phi}} \times \braket{\psi_{\theta,\phi}| U_{KI}[h]^{-t}|a_{3},b_{2},c_{2}} \nonumber \\
       \vdots \nonumber \\
    &\times  \braket{a_{2n},b_{1},c_{2n}| U_{KI}[h]^{t}|\psi_{\theta,\phi}} \times \braket{\psi_{\theta,\phi}| U_{KI}[h]^{-t}|a_{1},b_{2n},c_{2n}}.  \nonumber \\
    \end{align}
To evaluate the above expression we need to evaluate the following

\begin{align}
    \bra{ \mathbf{a},\mathbf{b},\mathbf{c}} (U_{KI} [\mathbf{h}])^{t})|\psi_{\theta,\phi} \rangle & = \sum_{s_{\tau}} \prod_{\tau=1}^{t-1} \braket{s_{\tau+1}|U_{KI}|s_{\tau}} \bra{a,b,c}U_{KI}\ket{s_{t}} \bra{s_{1}}\psi_{\theta,\phi}\rangle.
\end{align}
Using the following identities
\begin{align}
    \langle \mathbf{s} | \psi_{\theta,\phi}  \rangle &= \prod_{k=1}[\cos(\theta_{k}/2)\delta_{s_{k},1}+\sin(\theta_{k}/2)e^{i\phi_{k}}\delta_{s_{k},-1}], \\
    \braket{\mathbf{s} | U_{KI} | \mathbf{r} } &= \left( \frac{i}{2}\right)^{L/2} \exp[- \frac{i \pi}{4} \sum_{k=1}^{L}s_{k}r_{k}] \times  \exp\left[- \frac{i \pi}{4} \sum_{k=1}^{L}r_{k}r_{k+1}-i\sum_{k=1}^{L}h_{k}r_{k}\right], \\
    \braket{\mathbf{s} | e^{i(\pi/4)\sigma^{x}} | r } &= \sqrt{\frac{i}{2}}\exp\left[ -i \frac{\pi}{4} s r\right], \quad s,r \in\{\pm 1\},
\end{align}
we obtain
    \begin{align}
   \braket{ \mathbf{a},\mathbf{b},\mathbf{c}| (U_{KI} }[\mathbf{h}])^{t})|\psi_{\theta,\phi}  \rangle  &=  \left(\frac{i}{2}\right)^{(t L) / 2} \sum_{\left\{s_{\tau, k}\right\}} \exp \left[-\frac{i \pi}{4} \sum_{\tau=1}^{t} \sum_{k=1}^L s_{\tau, k} s_{\tau, k+1}-\frac{i \pi}{4} \sum_{\tau=1}^{t-1} \sum_{k=1}^L s_{\tau, k} s_{\tau+1, k}-i \sum_{\tau=1}^t \sum_{k=1}^L h_k s_{\tau, k}\right] \nonumber \\
& \times \exp \left[-\frac{i \pi}{4} \sum_{k=1}^{L_{A}} s_{t, k} a_k-\frac{i \pi}{4} \sum_{k=L_{A}+1}^{{L_{A}+L_{B}}} s_{t, k} b_{k-L_{A}}-\frac{i \pi}{4} \sum_{k=L_{A}+L_{B}+1}^{L} s_{t, k}\, c_{k-L_{A}-L_{B}}\right] \nonumber \\
&\times \left(\prod_{k=1}^{L}[\cos(\theta_{k}/2)\delta_{s_{1,k},1}+\sin(\theta_{k}/2)e^{i\phi_{k}}\delta_{s_{1,k},-1}] \right). \\
\end{align}
Combining above with Eq. \eqref{eq:evenmomsf} we obtain

    \begin{align}
&\tr((\rho_{AB}^{T_{B}})^{2n})=  \frac{1}{2^{2n L t}} \sum_{\left\{s_{\nu, \tau, k}\right\}} \exp \left[-i \sum_{k=1}^L \sum_{\nu=1}^{4 n} \text{sgn}(2n-\nu)\left(\sum_{\tau=1}^t\left(\frac{\pi}{4} s_{\nu, \tau, k} s_{\nu, \tau, k+1}+h_k s_{\nu, \tau, k}\right)+\sum_{\tau=1}^{t-1} \frac{\pi}{4} s_{\nu, \tau, k} s_{\nu, \tau+1, k}\right)\right] \nonumber \\
& \times \prod_{\nu=1}^{2n}\left\{\prod_{k=1}^{L_{A}}\left(1+s_{\nu, t, k} s_{2n+1+\bmod (\nu-2, 2n), t, k}\right) \prod_{k=L_{A}+1}^{L_{A}+L_{B}}\left(1+s_{\nu, t, k} s_{2n+1+\bmod (\nu, 2n), t, k}\right)\prod_{k=L_{A}+L_{B}+1}^{L}\left(1+s_{\nu, t, k} s_{2n+\nu, t, k}\right)\right\} \nonumber\\
& \times \prod_{\nu=1}^{4 n} \left(\prod_{k=1}^{L}[\cos(\theta_{k}/2)\delta_{s_{\nu,1,k},1}+\sin(\theta_{k}/2)e^{i\phi_{k}\mathrm{sgn}(2n-\nu)}\delta_{s_{\nu,1,k},-1}] \right).
\end{align}
Now, interpreting the total Hilbert space as $2n$ copies of $\mathcal{H}_{t}$ and noting the following matrix elements

\begin{align}
\left\langle\left\{s_{\nu, \tau}\right\}\right| \mathbb{T}^{(\theta,\phi)}_{C}[h]\left|\left\{r_{\nu, \tau}\right\}\right\rangle= & \frac{1}{2^{2(t-1) n}} \exp \left[-i \sum_{\nu=1}^{4 n} \operatorname{sgn}(2n-\nu)\left(\sum_{\tau=1}^t\left(\frac{\pi}{4} s_{\nu, \tau} r_{\nu, \tau}+h_j s_{\nu, \tau}\right)+\sum_{\tau=1}^{t-1} \frac{\pi}{4} s_{\nu, \tau} s_{\nu, \tau+1}\right)\right] \nonumber \\
& \times \prod_{\nu=1}^{2n}\left(\frac{1+s_{\nu, t} s_{\nu+2n, t}}{2}\right) \prod_{\nu=1}^{4 n} \left([\cos(\theta/2)\delta_{s_{\nu,1},1}+\sin(\theta/2)e^{i\phi \,\mathrm{sgn}(2n-\nu)}\delta_{s_{\nu,1},-1}] \right),
\end{align}\\ 

\begin{align}
\left\langle\left\{s_{\nu, \tau}\right\}\right| \mathbb{T}^{(\theta,\phi)}_{A}[h]\left|\left\{r_{\nu, \tau}\right\}\right\rangle= & \frac{1}{2^{2(t-1) n}} \exp \left[-i \sum_{\nu=1}^{4 n} \,\mathrm{sgn}(2n-\nu)\left(\sum_{\tau=1}^t\left(\frac{\pi}{4} s_{\nu, \tau} r_{\nu, \tau}+h_j s_{\nu, \tau}\right)+\sum_{\tau=1}^{t-1} \frac{\pi}{4} s_{\nu, \tau} s_{\nu, \tau+1}\right)\right] \nonumber \\
& \times \prod_{\nu=1}^{2n}\left(\frac{1+s_{\nu, t} s_{2n+1+\bmod (\nu-2, 2n), t}}{2}\right) \prod_{\nu=1}^{4 n} \left([\cos(\theta/2)\delta_{s_{\nu,1},1}+\sin(\theta/2)e^{i\phi \,\mathrm{sgn}(2n-\nu)}\delta_{s_{\nu,1},-1}] \right),\\
\left\langle\left\{s_{\nu, \tau}\right\}\right| \mathbb{T}^{(\theta,\phi)}_{B}[h]\left|\left\{r_{\nu, \tau}\right\}\right\rangle= & \frac{1}{2^{2(t-1) n}} \exp \left[-i \sum_{\nu=1}^{4 n} \,\mathrm{sgn}(2n-\nu)\left(\sum_{\tau=1}^t\left(\frac{\pi}{4} s_{\nu, \tau} r_{\nu, \tau}+h_j s_{\nu, \tau}\right)+\sum_{\tau=1}^{t-1} \frac{\pi}{4} s_{\nu, \tau} s_{\nu, \tau+1}\right)\right] \nonumber \\
& \times \prod_{\nu=1}^{2n}\left(\frac{1+s_{\nu, t} s_{2n+1\bmod (\nu, 2n), t}}{2}\right) \prod_{\nu=1}^{4 n} \left([\cos(\theta/2)\delta_{s_{\nu,1},1}+\sin(\theta/2)e^{i\phi \,\mathrm{sgn}(2n-\nu)}\delta_{s_{\nu,1},-1}] \right),
\end{align}
we obtain the following expression for the even moments, Eq \eqref{eq:evenmomsf}
  \begin{align}
   \tr((\rho_{AB}^{T_{B}})^{2n})=\tr\left[ \left(\prod_{k=1}^{L_{A}} \mathbb{T}^{(\theta,\phi)}_{A}[h_{k}]\right) \left( \prod_{k=L_{A}+1}^{L_{A}+L_{B}} \mathbb{T}^{(\theta,\phi)}_{B}[h_{k}]\right) \left(\prod_{k=L_{A}+L_{B}+1}^{L} \mathbb{T}^{(\theta,\phi)}_{C}[h_{k}]\right) \right].
\end{align}  
Note that $\mathbb{T}^{(\theta,\phi)}_{A}$ is a cyclic permutation of $\mathbb{T}^{(\theta,\phi)}_{C}$
\begin{equation}
    \mathbb{T}^{(\theta,\phi)}_{A}= \mathbb{P}\mathbb{T}^{(\theta,\phi)}_{C}\mathbb{P}^{\dagger},
\end{equation}
where $\mathbb{P}= \mathbb{1} \otimes \prod_{\nu=1}^{2n}\prod_{\tau=1}^{t}P_{(\nu,\tau),(\nu-1,\tau)}$ is the transposition operator with
\begin{equation}
    P_{(\nu,\tau),(\nu',\tau')} = \frac{1}{2}\mathbb{1}+ \frac{1}{2}\sum_{a\in\{x,y,z\}}\sigma^{a}_{\nu,\tau}\sigma^{a}_{\nu',\tau'},
\end{equation}
and the property that
\begin{equation*}
    \mathbb{P}^{-1}= \mathbb{P}^{\dagger}.
\end{equation*}
Similarly $\mathbb{T}_{B}$ is also a cyclic permutation of $\mathbb{T}^{(\theta,\phi)}_{C}$
\begin{equation}
    \mathbb{T}^{(\theta,\phi)}_{B}= \mathbb{P'}\mathbb{T}^{(\theta,\phi)}_{C}\mathbb{P'}^{\dagger},
\end{equation}
where $\mathbb{P'}= \mathbb{1} \otimes \prod_{\nu=1}^{2n}\prod_{\tau=1}^{t}P_{(\nu,\tau),(\nu+1,\tau)}$.

Thus, we finally have 
     \begin{align}\label{suppeq:tracefull}
   \tr((\rho_{AB}^{T_{B}})^{2n})=\tr\left[ \mathbb{P}\left(\prod_{k=1}^{L_{A}} \mathbb{T}^{(\theta,\phi)}_{C}[h_{k}]\right) \mathbb{P}^{\dagger} \mathbb{P'}\left( \prod_{k=L_{A}+1}^{L_{A}+L_{B}} \mathbb{T}^{(\theta,\phi)}_{C}[h_{k}] \right)\mathbb{P'}^{\dagger} \left(\prod_{k=L_{A}+L_{B}+1}^{L} \mathbb{T}^{(\theta,\phi)}_{C}[h_{k}]\right) \right].
\end{align}  

\section{Proof of Lemma 1 and Theorem 1}\label{sec:theolem1}

First, it is important to note that the transfer matrix for the negativity is exactly the transfer matrix obtained for R\'enyi entropy \cite{Bertini:2018fbz}. It is also shown in \cite{Bertini:2018fbz} that the solvability of any kind is obtained if we consider states belonging to $\mathcal{T}$ or $\mathcal{L}$ class, as given  by Eq.\eqref{suppeq:tclass} and \eqref{suppeq:lclass}. Specifically for the $\mathcal{T}$ class we have following properties \cite{Bertini:2018fbz}

\begin{prop*}
The spectrum of $\mathbb{T}^{(\pi/2,\phi)}_{C}[h]$ is fully determined as follows
\begin{enumerate}
    \item $\text{Spec}(\mathbb{T}^{(\pi/2,\phi)[h]}_{C}[h])= \{0,1\}$.
    \item The geometric multiplicity of the unit eigenvalue is one. 
\end{enumerate}
\end{prop*}
where $\text{Spec}(\bullet)$ denote the spectrum of $(\bullet)$.

Using results from \cite{Bertini:2018fbz} we also have the left eigenvector of $\mathbb{T}^{(\pi/2,\phi)}_{C}[h]$ as 
\begin{equation}
    \bra{\mathbb{1}}= \frac{1}{2^{nt/2}} \sum_{\{s_{\nu,t}\}} \bra{\{s_{\nu,t} \}} \otimes \bra{\{s_{\nu,t} \}},
\end{equation}
where $\bra{s_{\nu,t}}$ are the computational basis states.
We also note that properties of states belonging to $\mathcal{L}$ are same as that of $\mathcal{T}$ class, except the results are delayed by a period. 

Now taking the limit $L_{C} \rightarrow \infty$ we can rewrite Eq. \eqref{suppeq:tracefull} as follows
 \begin{align}\label{suppeq:negmom}
  \log( \tr((\rho_{AB}^{T_{B}})^{2n}))=\log(\braket{\mathbb{1}| \mathbb{P}\left(\prod_{k=1}^{L_{A}} \mathbb{T}_{C}[h_{k}]\right) \mathbb{P}^{\dagger} \mathbb{P'}\left( \prod_{k=L_{A}+1}^{L_{A}+L_{B}} \mathbb{T}_{C}[h_{k}] \right)\mathbb{P'}^{\dagger}|\mathbb{1}}).
\end{align} 
Then taking the thermodynamic limit, $L_{A} \rightarrow \infty$ and $L_{B} \rightarrow \infty$, 
 \begin{align}
\lim_{L_{A} \rightarrow \infty \atop L_{B} \rightarrow \infty }  \log( \tr((\rho_{AB}^{T_{B}})^{2n}))=  \log(\braket{|\mathbb{1}|\mathbb{P}|\mathbb{1}} \braket{|\mathbb{1}|\mathbb{P}^{\dagger}\mathbb{P'}|\mathbb{1}}\braket{|\mathbb{1}|\mathbb{P'}^{\dagger}|\mathbb{1}} ).
\end{align} 
Further noting that 
\begin{align}
    \braket{|\mathbb{1}|\mathbb{P}|\mathbb{1}}= \braket{|\mathbb{1}|\mathbb{P'}^{\dagger}|\mathbb{1}}&= 2^{(1-2n)t}, \\
    \braket{|\mathbb{1}|\mathbb{P}^{\dagger}\mathbb{P}'|\mathbb{1}} &= 2^{t(2-2n)t}.
\end{align}
Putting everything together we finally obtain
\begin{equation}
  \mathcal{E}_{2n}(t)=  \log( \tr((\rho_{AB}^{T_{B}}(t)^{2n})))= (4-6n)t\log(2).
\end{equation}
This thus prove Lemma 1.
Next, doing the analytic continuation $2n \rightarrow \alpha$ and then taking the limit $\alpha \rightarrow 1$, 
\begin{equation}
    \mathcal{E}(t)= \log(2)t.
\end{equation}
This also gives us the following for the spectrum of $(\rho_{AB}^{T_{B}}(t))$; $\textrm{Spec}(\rho_{AB}^{T_{B}}(t))$
\begin{equation}
    \textrm{Spec}(\rho_{AB}^{T_{B}}(t))= \{2^{-3t},0\},
\end{equation}
with degeneracy of non zero singular values as $2^{4t}$. By using the fact $\tr(\rho_{AB}^{T_{B}})=1$, the following relations are obtained
\begin{align}
    (N_{+}-N_{-}) &= 2^{3t}, \nonumber \\
    N_{+}+N_{-} &=2^{4t},
\end{align}
where $N_{\pm}$ denotes the number of positive and negative eigenvalues. Thus the number of negative and positive eigenvalues as 
\begin{equation}
    N_{+}= \frac{2^{4t}+2^{3t}}{2},\, N_{-}= \frac{2^{4t}-2^{3t}}{2},
\end{equation}
which proves the Theorem 1 of the main text. We emphasize that all the results obtained so far for the spectrum can be (and has been) easily verified for finite size systems as well.

\end{document}